\documentclass[11pt]{article}
\usepackage{amssymb, latexsym, amsmath}
\usepackage{amsfonts, graphics}
\usepackage{color} 
\usepackage{subfig}
\def\R{\mathbb R}

\def\be{\begin{equation}}
\def\ee{\end{equation}}
\def\bea{\begin{eqnarray}}
\def\eea{\end{eqnarray}}
\def\beas{\begin{eqnarray*}}
\def\eeas{\end{eqnarray*}}

\def\eps{\epsilon}

\begin{document}
\sloppy
\newtheorem{theorem}{Theorem}[section]
\newtheorem{definition}[theorem]{Definition}
\newtheorem{proposition}[theorem]{Proposition}
\newtheorem{example}[theorem]{Example}
\newtheorem{remark}[theorem]{Remark}
\newtheorem{cor}[theorem]{Corollary}
\newtheorem{lemma}[theorem]{Lemma}

\renewcommand{\theequation}{\arabic{section}.\arabic{equation}}

\title{Oscillating solutions of the Vlasov-Poisson 
       system---A numerical investigation}

\author{Tobias Ramming, Gerhard Rein\\
        Fakult\"at f\"ur Mathematik, Physik und Informatik\\
        Universit\"at Bayreuth\\
        D-95440 Bayreuth, Germany\\
        email: tobias.ramming@uni-bayreuth.de\\
        \phantom{ema}gerhard.rein@uni-bayreuth.de} 

\maketitle 

\begin{abstract}
Numerical evidence is given that spherically symmetric
perturbations of stable spherically symmetric steady states
of the gravitational Vlasov-Poisson system
lead to solutions which oscillate in time. The oscillations
can be periodic in time or damped.
Along one-parameter families of polytropic steady states
we establish 
an Eddington-Ritter type relation which relates
the period of the oscillation to the central density of the
steady state. The numerically obtained periods are used to
estimate possible periods for typical elliptical galaxies.
\end{abstract}
\section{Introduction}
\setcounter{equation}{0}

In astrophysics, a large ensemble of stars such as a galaxy or
a globular cluster is often modelled as a self-gravitating
collisionless gas which obeys the Vlasov-Poisson system
 \be \label{vlasov}
\partial_t f +v \cdot \nabla_x f - \nabla U \cdot \nabla _v f=0,
\ee
\be \label{poisson}
\Delta U = 4 \pi \rho,\ \lim_{|x|\to \infty} U(t,x)=0, 
\ee
\be \label{rhodef}
\rho (t,x) = \int f(t,x,v)\,dv. 
\ee
Here $f=f(t,x,v)\geq 0$ is the number density of the ensemble
in phase space and depends on time $t\in\R$, position $x\in\R^3$,
and velocity $v\in\R^3$, $\rho$ is the spatial mass density 
induced by $f$---unless explicitly stated otherwise integrals
always extend over $\R^3$---, and $U$ is the gravitational potential
generated by the ensemble. We assume that 
all the particles, i.e., stars, in the ensemble 
have the same mass which we normalize to unity. 
We refer to \cite{BT} for the astrophysics background of this system.

The initial value problem for the Vlasov-Poisson system is well
understood, and smooth initial data launch global smooth solutions,
cf.\ \cite{LP,Pf,Sch} or the review article \cite{Rein07}.
The system is known to have a plethora of steady states.
The spherically symmetric ones can be obtained by the following approach.
If the potential $U$ is time-independent and spherically symmetric,
then the particle energy and the angular momentum squared,
\be \label{parten}
E=E(x,v):=\frac{1}{2} |v|^2 + U(x),\  L := |x\times v|^2,
\ee
are constant along particle orbits, i.e., along
solutions of the characteristic system 
\[
\dot x = v,\ \dot v = - \nabla U(x)
\]
of the Vlasov equation (\ref{vlasov}). Hence an ansatz of the form
\be \label{ansatz}
f= \phi(E,L) 
\ee
satisfies the Vlasov equation and reduces the system to a
semilinear Poisson equation for $U$, which is obtained from (\ref{poisson})
by substituting
the ansatz into the definition (\ref{rhodef}) of the spatial density. 
The question which ansatz functions
$\phi$ lead to steady states which have
finite total mass and extension has been investigated by several
authors, and we refer to \cite{RR12} and the references there.
We will denote steady state quantities by $f_0$, $U_0$, etc.
Of obvious interest from the mathematics as well as applications point
of view is the nature of the dynamics in a neighborhood of such a steady state.
What is by now well understood is that these steady states are
stable provided $\phi$ is a strictly decreasing function of the energy
$E$ on the support of the steady state; for precise formulations
of such stability results we refer to \cite{G99,GR01,GR07,LMR,Rein07}
and the references there.

However, the fact that a particular steady state is stable does not
tell us the dynamical behavior of solutions which are launched by small
perturbations of it. This is the issue we study in the present paper
by numerical means. The somewhat surprising observation is that
all spherically symmetric and not too large perturbations of a
given stable steady state seem to launch solutions which oscillate in time
in the sense that their kinetic and potential energies oscillate and
their spatial support continues to expand and contract.
These oscillations can be time-periodic or damped, depending on the
steady state which is perturbed.
Such a behavior was already noticed in \cite{AR} for the
Einstein-Vlasov system. We have for the Vlasov-Poisson system
investigated these
oscillations with much higher numerical precision,
for much longer time spans and in a more systematic way.
We believe that they constitute an interesting new building block
for the picture of the overall dynamical possibilities of the
Vlasov-Poisson system, and to understand these time-periodic
oscillations in a mathematically rigorous way is
a challenging and worthwhile problem in mathematical physics, cf.\ \cite{HR}.

Our numerical results should be compared with 
the one-parameter family of semi-explicit
solutions constructed by {\sc Kurth} \cite{K}. Here
the spatial density is at each time constant on a ball
of radius $R(t)$ and zero elsewhere, and the free parameter
of the family is $\dot R(0)$. If $\dot R(0)=0$, the solution is a
steady state, and if $0<|\dot R (0)| <1$, then the function $R$
and hence the whole solution is time periodic with the amplitude
of the oscillations going to zero as $\dot R(0)$ goes to zero.
One can sum up the findings of our investigation by saying
that up to possible damping the Kurth family captures the generic picture of
the dynamics in a neighborhood of any stable steady state,
provided we restrict ourselves to spherical symmetry. 

The paper proceeds as follows. In the next section we 
reformulate the system in coordinates adapted to
spherical symmetry, and we discuss the type
of steady states which we perturb and the numerical approach 
used in our investigation.
In Section~\ref{sect-osc} we present our
numerical observations of oscillating solutions. 
We explain the various diagnostics
which were employed to numerically test the time-periodicity,
present some typical results which exhibit this property,
and we investigate the relation between the amplitude and
frequency of the oscillations. We find that in the limit
of small perturbations the frequency of the oscillations
does not depend on the type of perturbation but only on 
the perturbed steady state.
In Section~\ref{sect-edd} we consider one-parameter families of steady states
given by a fixed polytropic ansatz function and study the dependence of the 
period of oscillation
on the parameter which can be chosen as the central density of the steady state.
Such a relation is known for polytropic fluid models of stars. 
In the present context it
generalizes also to the unisotropic states.
In Section~\ref{sect-damp} we briefly consider the question for which
steady states the oscillations seem to be damped and for 
which they seem to be undamped and truly time periodic. We conclude with 
some final discussion of our observations in the last section.
\section{The spherically symmetric system and the numerical algorithm}
\label{sect-sphsym}
\setcounter{equation}{0}
By definition a distribution function $f$ is {\em spherically
symmetric} iff $f(t,x,v)= f(t,Ax,Av)$ for all rotations
$A\in \mathrm{SO}(3)$. By uniqueness, spherically symmetric initial
data launch spherically symmetric solutions. A spherically symmetric
distribution function can be written in the form
$f = f(t,r,w,L)$ where 
\[
r:=|x|,\ w:=\frac{x\cdot v}{r},\ L:= |x\times v|^2;
\]
$w$ is the radial velocity, and $L$, the modulus of angular momentum squared,
is conserved along particle trajectories due to spherical 
symmetry. In these variables the Vlasov-Poisson system takes the form
\be\label{vlasovss}
\partial_t f + w\, \partial_r f +\left(\frac{L}{r^3}-\partial_r U(t,r)\right)\, 
\partial_w f = 0,
\ee
\be \label{poissonss}
\partial_r U(t,r) = \frac{m(t,r)}{r^2},
\ee
\be \label{mdef}
m(t,r) = 4 \pi \int_0^r \rho(t,s)\, s^2 ds,
\ee
\be \label{rhoss}
\rho (t,r) = \frac{\pi}{r^2} 
\int_{-\infty}^\infty \int_0^\infty f(t,r,w,L)\, dL\, dw.
\ee
Here we have integrated the spherically symmetric Poisson equation once
and have put it into a form which is numerically easy to deal with.

As pointed out in the introduction there is a plethora of steady states
of the Vlasov-Poisson system. We restrict the general ansatz to the
technically convenient, more specific form
%:
\be \label{sansatz}
f= \phi(E_0-E)\,(L-L_0)_+^l. 
\ee
Here $l>-1/2$, $L_0 \geq 0$ is a cut-off angular momentum, $E_0 < 0$
is a cut-off energy, $\phi:\R \to [0,\infty[$ is measurable,
$\phi (\eta) = 0$ for $\eta < 0$, and
$\phi > 0$ a.~e.\ on $[0,\infty[$, 
and $(\cdot)_+$ denotes the positive part.
If $U_0=U_0 (r)$ denotes the potential of the spherically symmetric steady
state to be constructed then $y=E_0-U_0$ satisfies an equation of the
form
\be \label{yeq}
y' = - \frac{4 \pi}{r^2} \int_0^r s^{2l+2} g(y(s))\, ds
\ee
where $g\in C(\R)\cap C^1(]0,\infty[)$ is determined by $\phi$,
vanishes on $]-\infty,0]$, and is strictly positive on $]0,\infty[$.
For every prescribed value for $y(0)>0$ one obtains a unique solution of 
(\ref{yeq}) on $[0,\infty[$, and the corresponding steady state is 
compactly supported 
iff $y$ has a zero, i.e., $y(R)=0$ for some radius $R>0$. 
In \cite{RR12} and the references there one finds 
sufficient conditions on the ansatz function $\phi$ for this to happen.
The cut-off energy is then defined by $E_0=\lim_{r\to \infty} y(r)$.
In this way a fixed ansatz of the form  \eqref{sansatz}
leads to a one-parameter family of steady states which is parameterized
by $y(0)=E_0-U(0)$, the potential energy difference between the center and the 
spatial boundary of the state.

We mention examples of steady state ansatz functions
which play a role in our investigation:

\noindent
{\bf Polytropic balls.} Here
\be \label{poly}
f_0= (E_0-E)_+^k L^l,\ -1 < k < 3 l + 7/2.
\ee
In this case the steady state is supported on a ball of
radius $R$, and the steady state is isotropic or unisotropic depending on
whether $l=0$ or $l\neq 0$.

\noindent
{\bf Polytropic shells.} Here
\be \label{polysh}
f_0 = (E_0-E)_+^k (L-L_0)^l,\ -1 < k < 3 l + 7/2,\ L_0>0.
\ee
In this case the steady state is supported on a shell
with inner radius $R_i = \sqrt{L_0/(2 y(0))}$ and outer radius $R>R_i$,
and the steady state is unisotropic.

\noindent
{\bf King's model.} Here
\be \label{king}
f_0 = (e^{E_0 - E}-1)_+.
\ee
In this case the steady state is supported on a ball
of radius $R$ and isotropic. All these steady states have
finite mass and compact support, and they are known to be stable.

\noindent
{\bf Kurth's model.}
Here
%\[
\be \label{kurth}
f_0(x,v) = \frac{3}{4\pi^3}
\left\{
\begin{array}{cl} 
\left(1-|x|^2 - |v|^2 + L\right)^{-1/2}&,\
\mbox{where}\ (\ldots)>0 \ \mbox{and}\ L<1,\\
0&,\ \mbox{else}
\end{array} \right.
\ee
%\]
defines a steady state with spatial density and potential
\[
\rho_0(x) = \frac{3}{4\pi} \mathbf{1}_{B_1}(x),\ \ 
U_0(x) =
\left\{
\begin{array}{cl}
|x|^2 /2 - 3/2&,\ |x|\leq 1,\\
-1/|x| &,\ |x|>1;
\end{array} \right.
\]
notice that $f_0$ is again a function of $E$ and $L$.
The importance of this model for the present investigation
lies in the fact that
the transformation
\[
f(t,x,v) = f_0(x/R(t),R(t)v - \dot R(t) x)
\]
turns this steady state into a time dependent solution
with spatial mass density 
\[
\rho(t) = \frac{3}{4\pi} \frac{1}{R^3(t)} \mathbf{1}_{B_{R(t)}}, 
\]
provided the function $R=R(t)$ solves the differential equation
\[
\ddot R - R^{-3} + R^{-2} = 0,
\]
and $R(0)=1$. The only free parameter in this family is $\eps = \dot R(0)$.
For $\eps = 0$ one recovers the steady state $f_0$, and for $0<|\eps|<1$
the function $R$ and hence the solution $f$, which for $|\eps|$ small
is a small perturbation of $f_0$, is time periodic with period
$2\pi (1-\eps^2)^{-3/2}$.

In our numerical simulations we used the following types of perturbations
of such steady states:
\begin{itemize}
\item[(P1)]
Perturbations by amplitude: $\mathring f=(1+\eps) f_0$ with $\eps \in \R$ small.
\item[(P2)]
Perturbations by shift: $\mathring f = f_0(\cdot + (r,w,L)_\eps)$ 
where the displacement vanishes as $\eps \to 0$.
\item[(P3)]
Kurth-type perturbations: $\mathring f(x,v) = f_0(x, v - \eps x)$.
\item[(P4)]
Dynamically accessible perturbations:
During an initial time interval $[0,t_{\mathrm{pert}}]$
the initial data
$f_0$ is evolved under the influence of an external
field $\eps F$ or under the modified self-consistent
field $-(1+\eps) \nabla U_0$, and the perturbation $\mathring f$
is defined as this evolved state evaluated at $t=t_{\mathrm{pert}}$.
\end{itemize}
Perturbations of the latter type  preserve all the Casimir invariants
$\int C(f)\,dv\,dx$ of the Vlasov-Poisson system.

For the numerical simulations we used a particle-in-cell scheme
which we briefly review in the spherically symmetric set-up.
Given spherically symmetric and compactly supported initial data 
$\mathring f$ we initialize the scheme
by splitting the support into a finite number of disjoint cells.
Into each cell we place a numerical particle with a weight given
by the volume of that cell times the value 
of $\mathring f$ at the position of the particle.
In the time step we take such a collection of particle
positions and corresponding weights and compute from this an approximation
of $\rho$ and hence of $m$ and $\partial_r U$ on a grid
in the radial direction. This grid is chosen such that it 
covers the $r$-interval in which the numerical
particles are currently found. Using the approximation of 
$\partial_r U$ the particle positions can now be updated by moving
them according to the characteristic system of the Vlasov equation 
(\ref{vlasovss}):
\be \label{cs}
\dot r = w,\ \dot w = \frac{L}{r^3}-\partial_r U(t,r),\ \dot L = 0;
\ee
close to the origin it is advantageous to use Cartesian coordinates
for propagating the particles. We found that a simple Euler scheme is 
sufficient for propagating the particles. 
The weights of the particles are not changed in the time step,
which reflects the fact that the characteristic flow conserves
phase space volume.

The conservation of the total mass
$\iint f(t,x,v)\,dv\,dx$  is a generic feature of 
this scheme. On the other hand, solutions of the Vlasov-Poisson system
also conserve energy ${\cal H} = E_\mathrm{kin} + E_\mathrm{pot}$,
the kinetic and potential parts of which are defined by
\[
E_\mathrm{kin}(t) = \frac{1}{2} \iint |v|^2 f(t,x,v)\, dv\, dx,\qquad  
E_\mathrm{pot}(t) = 
-\frac{1}{8\pi} \int |\nabla U(t,x)|^2\, dx.
\]
Conservation of energy is not built into the scheme and can 
hence be used to monitor its accuracy.

The above scheme is easy to parallelize. Each processor is responsible
for a fixed batch of the numerical particles, i.e., for propagating them
according to (\ref{cs}) and for computing their contribution to
$\rho$. Of course at each time step these contributions have to be added
up to obtain the total spatial density $\rho$ and the induced field according to
(\ref{poissonss}), but since the number of grid points
of the spatial grid in the radial direction is very much smaller than 
the number of numerical particles, the scheme scales very well when
the number of processors is increased, cf.~\cite{KRR}.
\section{Oscillating solutions}
\label{sect-osc}
\begin{figure}
\centering
\vspace{-.5cm}
\subfloat[Spatial density]
{\resizebox*{!}{0.44\textwidth}{\input{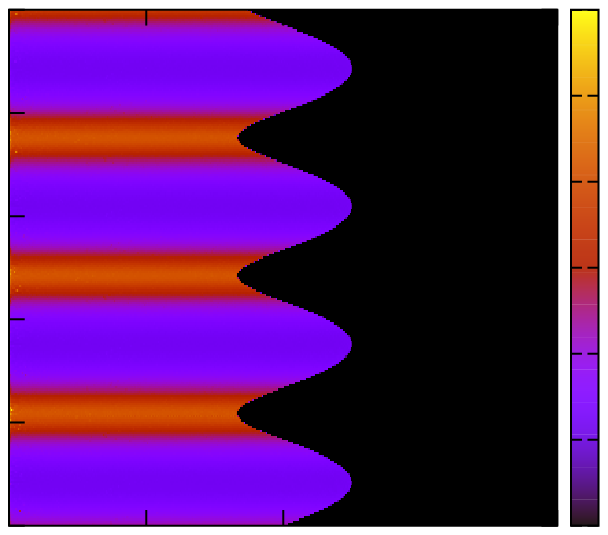}}}
\hfill
\subfloat[Potential energy]
{\resizebox*{!}{0.44\textwidth}{\input{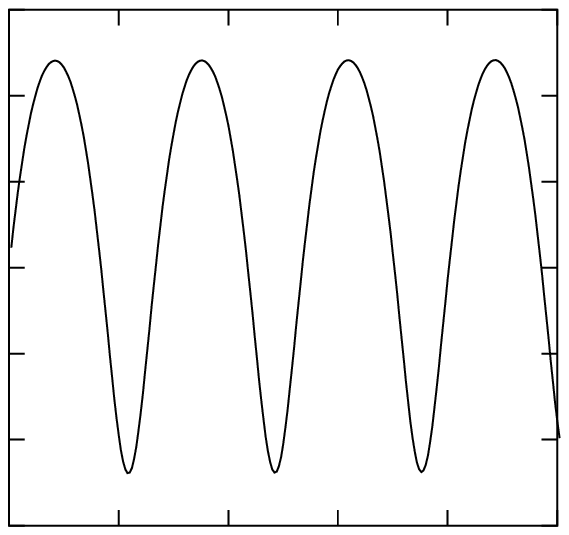}}}
\caption{Time evolution of the Kurth solution with 
$\eps=0.2$. The calculation used about $36\cdot 10^6$ 
particles% ($dr=0.004$, $du=0.003$, $d\alpha=0.006$)
.}
\label{plot_kurth_02}
\end{figure}

\setcounter{equation}{0}
Since we want to numerically investigate the question whether or not 
small perturbations of steady states lead to oscillating 
behavior, it seems worthwhile to first test
the employed particle-in-cell code
on the family of Kurth solutions \eqref{kurth}.
As mentioned before, for $0<\eps<1$ these solutions are
time-periodic with period 
$2\pi (1-\eps^2)^{-3/2}$. The numerically computed
density profile of the solution 
for $\eps=0.2$ is presented in Figure~\ref{plot_kurth_02}(a); 
its potential energy as a function of time is shown 
in Figure~\ref{plot_kurth_02}(b). 
Table \ref{table_kurth_periods} shows the dependence 
of the period on the parameter $\eps$ together
with some corresponding numerical results. These 
match the theoretical predictions 
quite well. Also the discontinuity of the spatial density 
$\rho$ at $|x|=R(t)$ is recovered nicely. Given the fact that 
Lagrangian methods based on the strong formulation 
of the problem suffer seriously from low regularity
and that for the Kurth solutions
$\rho$ is discontinuous and $f$ singular at
the boundaries of their respective supports,
the code passes this test quite well;
the solutions for which we want to conclude an oscillatory, time-periodic 
behavior from the numerical simulations are much smoother
than the Kurth solutions.

%We see a slightly larger deviation 
%from the theoretical prediction for $\eps=0.4$; it 
%stems from the singularity of the phase 
%space density of the kurth solutions at the boundary of 
%the support. Increasing the number of particles used for 
%the calculation from approximately $4.5m$ to $8m$ yields 
%a new period length of $8.10$. \textcolor{blue}{Es laufen 
%gerade noch Rechnungen mit einer 
%deutlich höheren Auflösung - wie das Beispiel für 
%$\eps=0.2$; diese zeigen was die Massen 
%angeht nur noch Abweichungen von $<5.0e-4$. Der ganze 
%Abschnitt sollte sich also erledigen... hoffentlich...}

\begin{figure}
\centering
\vspace{-.5cm}
\subfloat[$\Delta_\rho$]
{\resizebox*{!}{0.455\textwidth}{\input{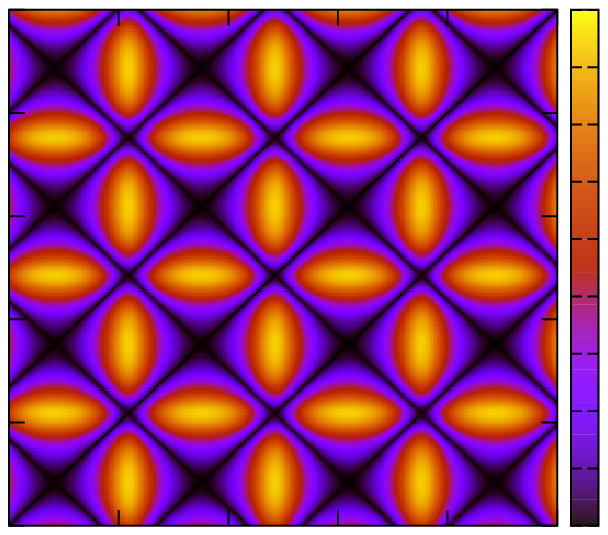}}}
\hfill
\subfloat[$\Delta_f$]
{\resizebox*{!}{0.455\textwidth}{\input{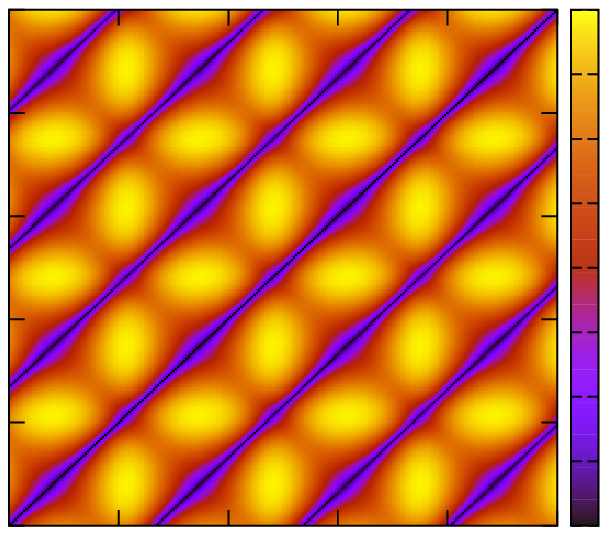}}}
\caption{$\Delta_\rho$ and $\Delta_f$ for to the Kurth solution 
with $\eps=0.2$. 
The calculation used about $36\cdot10^6$ particles.}
\label{plot_kurth_02_L1diff}
\end{figure}

Knowing that Kurth solutions are periodic in time, it is 
easy to determine their period from 
the previously shown results, but it is more 
involved to decide whether some periodic looking 
numerical result originates from a true periodicity of the 
underlying solution.
To do so, we must in principle
investigate the periodicity of the particle distribution function
$f$, since the latter determines the dynamics and not
some derived, macroscopic quantity like $\rho$ or the 
kinetic or potential energy. 
To this end we use the fact that a continuous function 
$\phi \colon \R\times\R^n \to \R$ is periodic in time 
with period $T>0$, if it holds that 
\[
\|\phi(t,\cdot)-\phi(s,\cdot)\|_{L^1(\R^n)}=0, 
\quad s=t+kT,\; k\in\mathbb Z,\; t\in\R.
\]
\begin{table}[ht]
\centering
\begin{tabular}{|c|c|c|c|c|c|} \hline
$\eps$ &  $0.1$ & $0.2$ & $0.3$ & $0.4$ & $0.5$ \\ \hline
$\lambda_e$ & $6.38$ & $6.68$ & $7.24$ & $8.16$ & $9.67$ \\
$\lambda_n$ & $6.39$ & $6.70$ & $7.24$ & $8.16$ & $9.66$\\ \hline
\end{tabular}
\caption{Exact $\lambda_e$ and numerical $\lambda_n$ 
periods of the Kurth solutions. The calculation used about 
$33\cdot 10^6-36\cdot 10^6$ particles.}
\label{table_kurth_periods}
\end{table}
We approximate the above norm by its discrete 
$L^{1,h}$-version, based on the particle discretisation 
used in our PIC code, and since we only have 
information on our approximation at a discrete set 
of times $\mathcal T := \{t_j\}$, we investigate the 
function
\[
\Delta_f \colon {\mathcal T}^2\rightarrow\R,\quad 
(t_i,t_j)\mapsto\|f(t_i,\cdot)-f(t_j,\cdot)\|_{L^{1,h}(\R^6)};
\]
the function $\Delta_\rho$ is defined  in terms of $\rho$ in the analogous way.
The results are shown in Figures~\ref{plot_kurth_02_L1diff}~(a) and (b).
We see that there is a time $T$ such that when $t_i-t_j=k T$
for some integer $k$ then
both $\Delta_f(t_i,t_j)$ and $\Delta_\rho(t_i,t_j)$ become (almost)
zero which indicates periodicity with period $T$. Notice that
$\Delta_\rho(t_i,t_j)$ also becomes (almost) zero 
when $t_i + t_j=k T + s$ for some fixed $s$. This is due to the 
fact that $\rho(T_0 + t) = \rho(T_0 - t)$ where $T_0$ is one of the times
when the state has maximal (or minimal) extension and $t$ is arbitrary,
cf.\ Figure~\ref{plot_kurth_02}~(a).

\begin{figure}
\centering
\vspace{-.65cm}
\subfloat[Spatial density]
{\resizebox*{!}{0.46\textwidth}
{\input{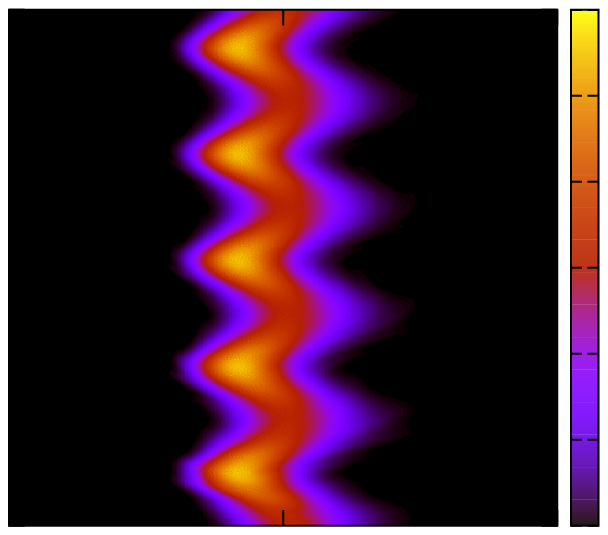}}}
\hfill
\subfloat[Potential energy]
{\resizebox*{!}{0.46\textwidth}
{\input{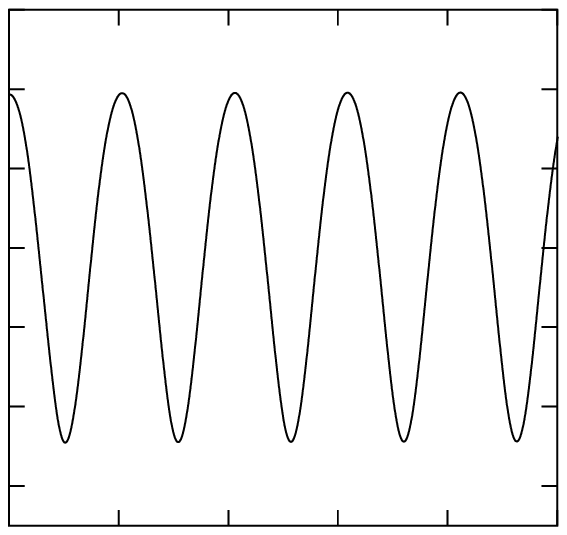}}}\\
\vspace{-.85cm}
\subfloat[Spatial density]
{\resizebox*{!}{0.46\textwidth}
{\input{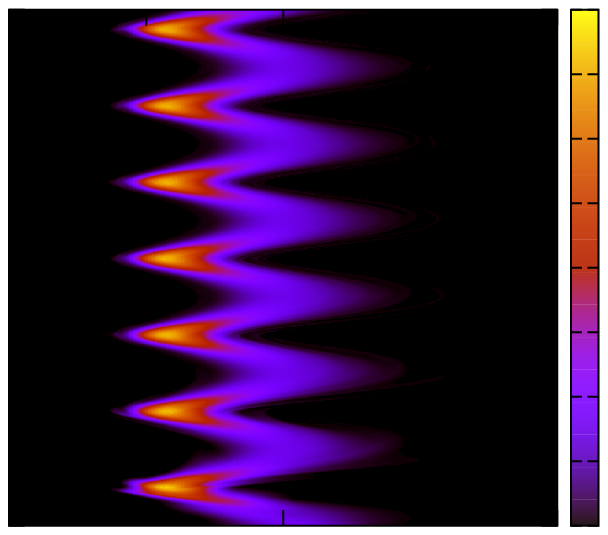}}}
\hfill
\subfloat[Potential energy]
{\resizebox*{!}{0.46\textwidth}
{\input{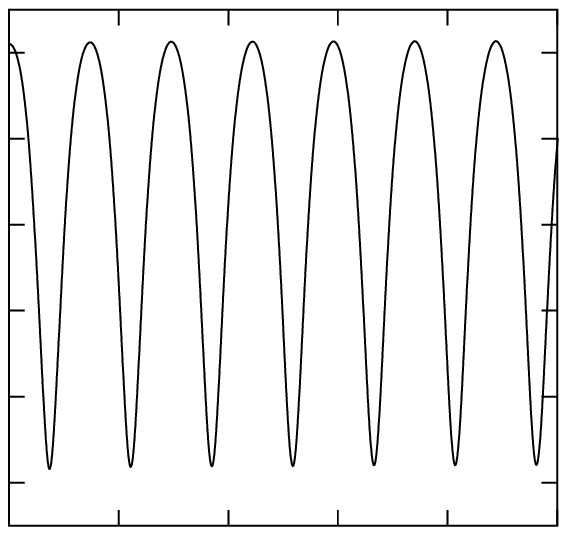}}}\\
 \vspace{-.0cm}
\caption{Shell for $k=1.0$, $l=0.5$ and $L_0=1.0$ 
after a Kurth-type perturbation with $\eps=0.05$ (upper) 
and $\eps=0.2$ (lower) respectively.}
\label{plot_kurth_005_02}
\end{figure}

\begin{figure}
\centering
\vspace{-.65cm}
\subfloat[Spatial density]
{\resizebox*{!}{0.46\textwidth}{\input{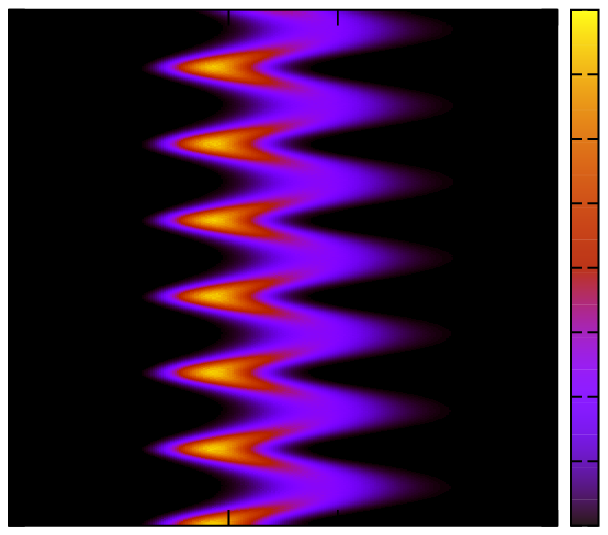}}}
\hfill
\subfloat[Potential energy]
{\resizebox*{!}{0.46\textwidth}{\input{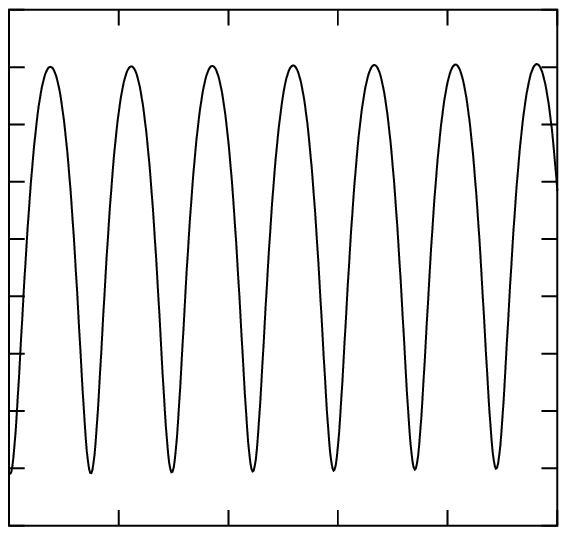}}}\\
\vspace{-.85cm}
\subfloat[$\Delta_\rho$]
{\resizebox*{!}{0.46\textwidth}{\input{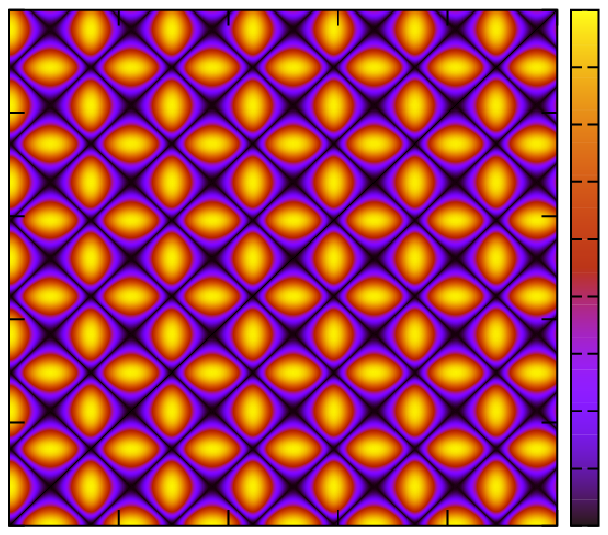}}}
\hfill
\subfloat[$\Delta_f$]
{\resizebox*{!}{0.46\textwidth}{\input{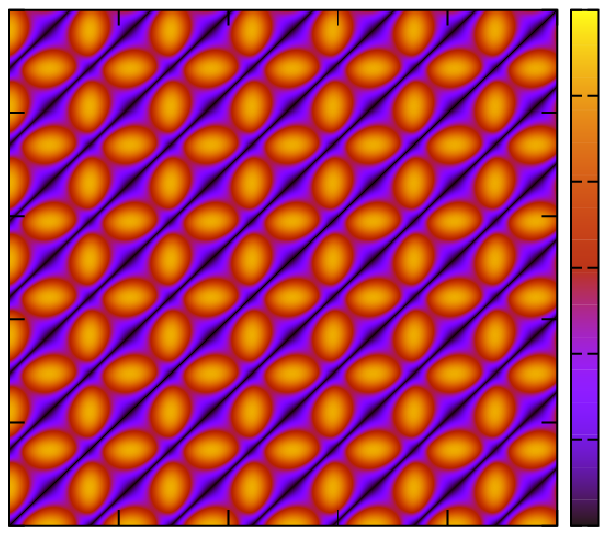}}}\\
 \vspace{-.0cm}
\caption{Shell solution with a dynamically accessible perturbation.}
\label{plot_shell_DA}
\end{figure}

We now discuss the results
obtained by perturbing the steady states mentioned above. 
The first thing one may ask is whether a
Kurth-type perturbation applied to a different steady state 
again leads to a periodic solution. 
To begin with, we consider the polytropic shell with 
$k=1.0$, $l=0.5$, $L_0=1.0$, and $y(0)=1.0$. 
Figure~\ref{plot_kurth_005_02} shows the solutions triggered
by Kurth-type perturbations with $\eps=0.05$ and $\eps=0.2$,
which appear to be (very close to) periodic.
It is natural to ask, whether perturbations of different 
types launch solutions with a quasi-periodic behavior as well. It
turns out that this is the case for most combinations 
of the perturbations and steady states mentioned above. 
Figure~\ref{plot_shell_DA} shows a dynamically accessible 
perturbation of a shell solution, computed on a somewhat 
longer time interval. The figure also contains the potential energy and 
the functions $\Delta_\rho$ and $\Delta_f$ related to the solution.

\begin{figure}
\centering
\vspace{-.5cm}
\subfloat[King-type solutions]
{\resizebox*{!}{0.455\textwidth}{\input{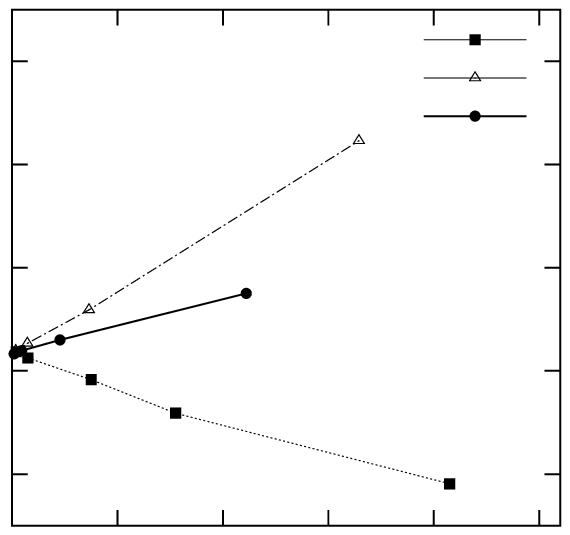}}}
\hfill
\subfloat[Shells]
{\resizebox*{!}{0.455\textwidth}{\input{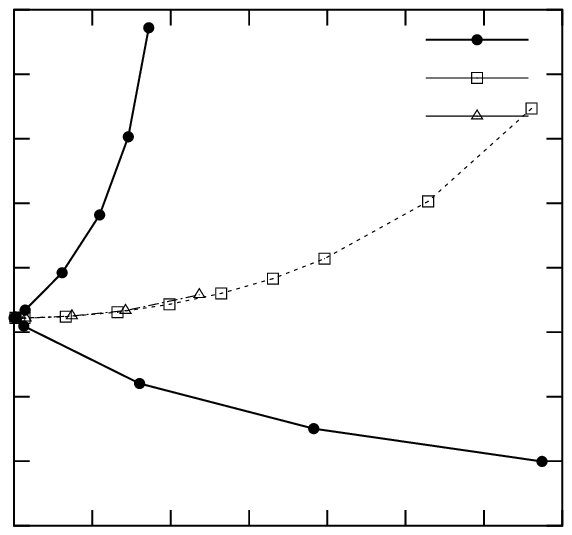}}}
\caption{Relation of period and amplitude for different kinds of perturbations.} 
\label{plot_perioden}
\end{figure}

It is worth mentioning that for a given steady state all the perturbations 
we applied to it led to oscillations of the same type. In particular,
if the intensity of the perturbation, i.e., the amplitude of 
the oscillation,  is reduced, then the period converges to a value which
depends only on the specific steady state and not on the 
type of the perturbation.
We were not able to find a clear numerical indication
of different modes of the oscillations, but in Section~\ref{sect-damp}
we report some observations which may be interpreted as superpositions of
two different modes, cf.\ 
Figures~\ref{damping2} and \ref{damping3}.
Figure~\ref{plot_perioden} shows our results for one specific King and one 
specific shell solution.
\section{The Eddington-Ritter relation}
\label{sect-edd}
\setcounter{equation}{0}
The light variations of the Cepheid variables 
can be explained by modelling them as
time periodic pulsations of solutions to the Euler-Poisson 
system---for a review of the corresponding history
we refer to \cite{ross}.
Linearizing the Euler-Poisson system about a polytropic steady state
Eddington \cite{edd} derived the following relation between the central density 
$\rho(0)$ of the steady state and the period $T$ of an oscillatory solution
of the linearized system, 
triggered by a small perturbation of the 
steady state: $\rho(0)^{1/2} T = {const}$, where the constant
depends on the parameters of the polytropic equation of state.
Since such a relation was earlier suggested by Ritter, 
we refer to
it as the {\em Eddington-Ritter relation}.

Isotropic steady states of the Vlasov-Poisson system
are in one-to-one correspondence with those of the Euler-Poisson system
with a suitable equation of state. For an isotropic state
of the form \eqref{poly} with $l=0$
the relation $\rho(0) = c_k y(0)^{k+3/2}$ holds; we recall
that a fixed ansatz function---in the present case fixed
values for $k$ and $l$---gives rise to a one-parameter
family of steady states parameterized by $y(0)=E_0 - U(0)$.
It is tempting
to fix some $k$ in \eqref{poly} and determine numerically the quantity
$y(0)^{k/2+3/4} T$ for different choices of $y(0)$; $T$ is the period
of the oscillation.
\begin{table}[ht]
\centering
\begin{tabular}{|c|c|c|} \hline
$y(0)$ & $T$ & $c=y(0)^{3/4} T$ \\ \hline
$0.6$  & $ 1.943 $ &  $ 1.32451 $\\
$0.8$  & $ 1.569 $ &  $ 1.32758 $\\
$1.0$  & $ 1.332 $ &  $ 1.33214 $\\
$1.2$  & $ 1.167 $ &  $ 1.33800 $\\
$1.4$  & $ 1.044 $ &  $ 1.34364 $\\
$1.6$  & $ 0.947 $ &  $ 1.34794 $\\
\hline
\end{tabular}
\caption{Eddington-Ritter relation for $k=0$ and $l=0$}
\label{tablek0}
\end{table}
In Table~\ref{tablek0} and Table~\ref{tablek1} this is done
for $k=0$ and $k=1$ respectively. It should be noted that for $k=0$
the ratio of the maximal and minimal value for $T$ equals
$T_\mathrm{max}/T_\mathrm{min} = 2.051$ while the ratio of the corresponding ``constants''
is $c_\mathrm{max}/c_\mathrm{min} = 1.0177$;
for $k=1$ we find
$T_\mathrm{max}/T_\mathrm{min} = 3.400$ and $c_\mathrm{max}/c_\mathrm{min} = 1.003$.
Analogous result were found for different values of $k$.
Hence one may indeed claim that an Eddington-Ritter relation
does hold for the oscillations of galaxies. We should emphasize at this
point that the periods in the tables above are obtained from
simulations of oscillating solutions to the fully non-linear 
Vlasov-Poisson system which are triggered by small perturbations of 
the given steady state.
\begin{table}[ht]
\centering
\begin{tabular}{|c|c|c|} \hline
$y(0)$ & $T$ & $c=y(0)^{1/2+3/4} T$ \\ \hline
$0.6$  & $ 5.233 $ &  $ 2.76311 $\\
$0.8$  & $ 3.650 $ &  $ 2.76157 $\\
$1.0$  & $ 2.761 $ &  $ 2.76111 $\\
$1.2$  & $ 2.200 $ &  $ 2.76312 $\\
$1.4$  & $ 1.815 $ &  $ 2.76399 $\\
$1.6$  & $ 1.539 $ &  $ 2.76922 $\\
\hline
\end{tabular}
\caption{Eddington-Ritter relation for $k=1$ and $l=0$}
\label{tablek1}
\end{table}
For non-isotropic steady states of the Vlasov-Poisson system,
for example states of the form \eqref{poly} with $l\neq 0$,
there do not exist corresponding steady states of
the Euler-Poisson system. It is therefore an interesting question
whether an Eddington-Ritter-type relation still holds for
non-isotropic states, say, for states of the form \eqref{poly}
with $l\neq 0$. It is a-priori not obvious what the corresponding 
relation might be. In \cite{HR} the authors aim for a mathematical 
analysis of the oscillatory solutions to the Vlasov-Poisson
system, and in the course of this investigation the relation
\begin{equation} \label{hrr_rel}
y(0)^{\frac{k+2l+3/2}{2l+2}} T = const
\end{equation}
was formally derived, where the right hand side depends on $k$ and $l$.
We first observe that this relation reduces to the one which we checked 
numerically for the isotropic case $l=0$. The results
of a corresponding numerical check for the non-isotropic case
$k=0,\ l=2$ are given in Table~\ref{tablek0l2}.
\begin{table}[ht]
\centering
\begin{tabular}{|c|c|c|} \hline
$y(0)$ & $T$ & $c=y(0)^{\frac{4+3/2}{4+2}} T$ \\ \hline
$0.6$  & $ 6.650 $ &  $ 4.16352 $\\
$0.8$  & $ 5.113 $ &  $ 4.16677 $\\
$1.0$  & $ 4.167 $ &  $ 4.16731 $\\
$1.2$  & $ 3.529 $ &  $ 4.17115 $\\
$1.4$  & $ 3.066 $ &  $ 4.17320 $\\
$1.6$  & $ 2.717 $ &  $ 4.17972 $\\
\hline
\end{tabular}
\caption{Eddington-Ritter relation for $k=0$ and $l=2$}
\label{tablek0l2}
\end{table}
In this case
$T_\mathrm{max}/T_\mathrm{min} = 2.448$ while
$c_\mathrm{max}/c_\mathrm{min} = 1.0039$. 

In all the above numerical runs the system was evolved up to time $50$
in $3\cdot 10^5$ time steps, using between $19\cdot 10^6$ and $37 \cdot 10^6$
particles and dynamically accessible perturbations.
The period was then determined from the fluctuations in the kinetic energy. 

It seems fair to say that we have 
numerically verified (and in part analytically derived) an
extension of the Eddigton-Ritter relation to steady states 
of the Vlasov-Poisson system of polytropic form \eqref{poly},
including the non-isotropic case for which such a relation has no analogue
in the fluid case.
\section{Damping or no damping}
\label{sect-damp}
\setcounter{equation}{0}

In the above
we have restricted ourselves to the gravitational
case of the Vlasov-Poisson system. If the sign in the right hand side of 
the Poisson equation (\ref{poisson}) is reversed, one obtains the
plasma physics version of the system. 
In this case
one can add a fixed, spatially homogeneous ion background,
which then allows for spatially homogeneous and
pointwise neutral steady states with vanishing electrostatic field. 
Based on an analysis by linearization 
{\sc L.~Landau} in 1946 predicted the phenomenon which was 
later termed ``Landau damping'',
whereby for small perturbations of these homogeneous equilibria 
the electrostatic field damps out to zero \cite{L}. 
In the celebrated paper \cite{MV}
the phenomenon of Landau damping
was rigorously established on the non-linear level. 
In the gravitational case at hand
the situation
is more complicated, in particular, no spatially homogeneous steady states
exist except for the vacuum state. Nevertheless,
similar damping phenomena are discussed in
the astrophysics literature, cf.\ \cite{BT} and the references there or in 
\cite{MV}. 
The explicit solution family of 
{\sc Kurth},
where exactly time periodic solutions are 
launched by small perturbations of a stationary one,
shows that Landau damping need not always occur
in the gravitational case.

Following \cite{MV}, three ingredients seem important for
Landau damping to happen: The perturbed steady state should be rather smooth,
it should be spatially homogeneous or at least close to homogeneous on
part of its support, and the perturbation should be sufficiently small.

\begin{figure}
\centering
\vspace{-.65cm}
\subfloat[$k=0.5,\ l=0$]
{\resizebox*{!}{0.46\textwidth}{\input{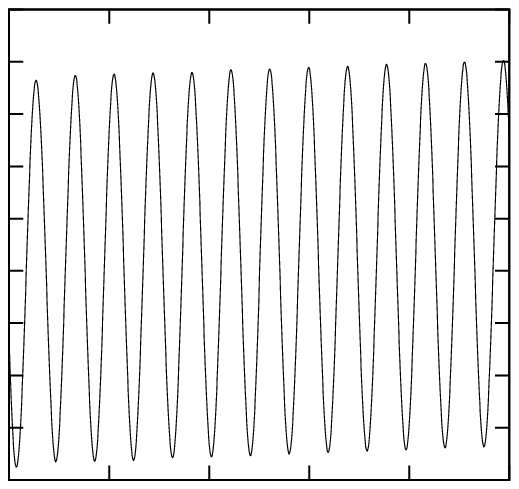}}}
\hfill
\subfloat[$k=1.2,\ l=0$]
{\resizebox*{!}{0.46\textwidth}{\input{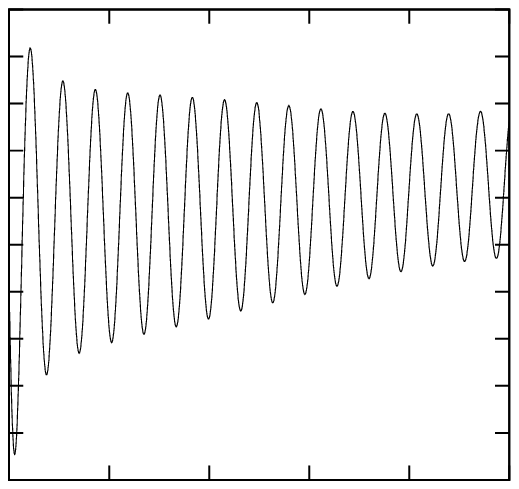}}}\\
\vspace{-.85cm}
\subfloat[$k=1.6,\ l=0$]
{\resizebox*{!}{0.46\textwidth}{\input{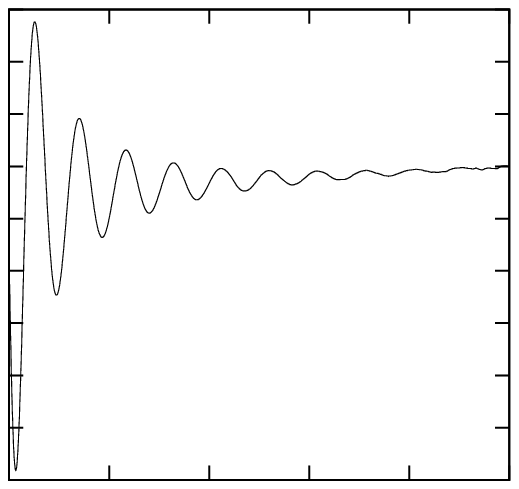}}}
\hfill
\subfloat[$k=3.0,\ l=5.0$]
{\resizebox*{!}{0.46\textwidth}{\input{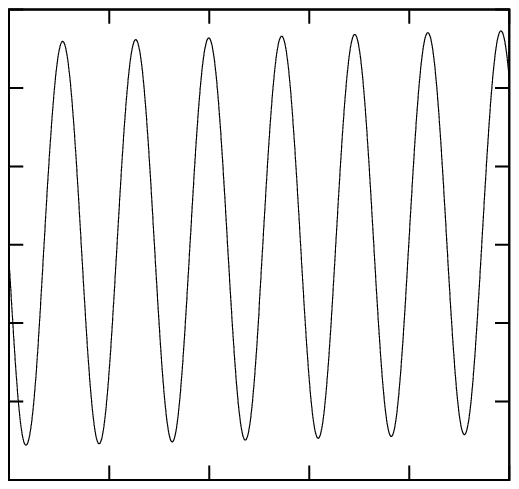}}}\\
 \vspace{-.0cm}
\caption{Damping phenomena.}
\label{damping1}
\end{figure}

In Figure~\ref{damping1} we
plot the potential energy
for a perturbation of a polytropic steady state for four 
different choices of $k$ and $l$;
the perturbation is dynamically accessible, and about
$20\cdot 10^6$ particles have been used.
The first three plots show that
with $l=0$ and increasing $k$, which corresponds to
increasing smoothness of the steady state, the amplitude of the oscillation
becomes more strongly damped. However, smoothness of the perturbed steady state
alone is not sufficient for damping to take place. 
In Figure~\ref{damping1}~(d)
we perturbe a polytropic state with $k=3$ and $l=5$, which is smoother
than the previous ones, but no damping is observed. A possible explanation
is the fact that for $l=0$ the polytropic steady states
have a strictly decreasing spatial density with $\rho' (0)=0$,
whereas $l>0$ leads to steady states with $\rho(0)=0$,
where $\rho$  strictly increases up to some maximum and 
than strictly descreases.
\begin{figure}
\centering
\resizebox*{!}{0.6\textwidth}{\input{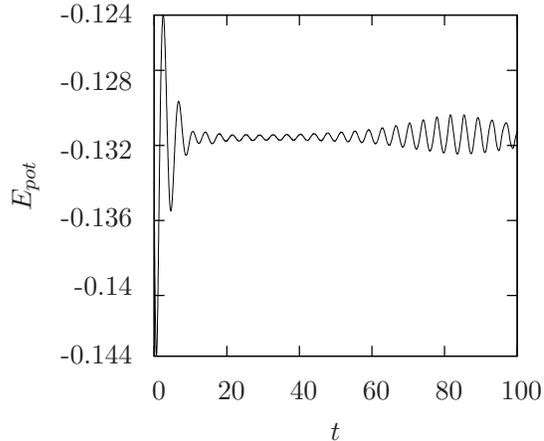}}
\caption{$k=1.6,\ l=0$ with a stronger perturbation.}
\label{damping2}
\end{figure}
The latter behavior can be considered as more strongly inhomogeneous than 
the former. This also fits with the observation that no damping seems to occur
for polytropic shells. It should be noted that in the examples above the
perturbation is always rather small, as can be seen from the fact that
initially the potential energy deviated only very little from is mean value. 

In Figure~\ref{damping2} we again consider the 
fairly smooth isotropic polytrope
$k=1.6$ and $l=0$, but we consider a stronger perturbation.
Had we stopped the computation at $T=20$, this might have been 
tabbed as another example of a damped oscillation, but 
actually something quite different (and quite a bit more interesting)
happens. A possible interpretation of this phenomenon
is that we here observe a superposition of two different oscillatory modes.
\begin{figure}
\centering
\resizebox*{!}{0.6\textwidth}{\input{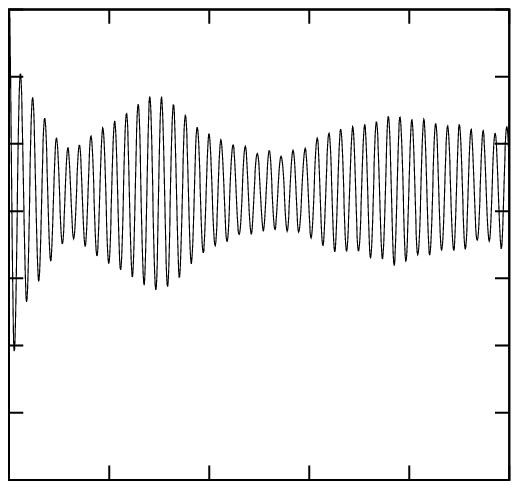}}
\caption{The King model with a stronger perturbation.}
\label{damping3}
\end{figure}
In Figure~\ref{damping3} we see a similar effect as in Figure~\ref{damping2}
for a perturbation of the King model, but we have so far not obtained a 
clear picture of the latter phenomenon, nor indeed of 
when oscillations are damped 
and when they are not.
\section{Final comments}
\label{sect-findisc  }
\setcounter{equation}{0}
One should ask whether our numerical observations
exhibit a genuine feature of the dynamics of the Vlasov-Poisson system
or whether they are a numerical artefact. Firstly, it is easy to imagine
that numerical effects destroy features like periodic orbits, but is seems
hard to imagine that they generate these features. More importantly,
for the Euler-Poisson system it is known by rigorous analysis that
such time periodic oscillations exist on the linearized level
and that in a well-defined sense they approximately survive 
for the non-linear system,
cf.\ \cite{edd,jang,mak,ross}.
The fact that we recover the Eddington-Ritter relation, which is 
known in the Euler-Poisson context, also in the Vlasov-Poisson context and that
in \cite{HR} a formal derivation of this relation from a suitable
linearized system is obtained
is a strong indication that
the observed oscillations are a genuine feature of the Vlasov-Poisson dynamics.

Finally, one should ask whether our numerical observations
are relevant for astrophysics. In this context it seems of interest to
express the observed periods in suitable units.
By matching the numerically obtained values for the radius and mass
of a given steady state with observed data for a real galaxy
we fix the units of length and mass. Moreover, the 
Vlasov-Poisson system contains only one physical constant,
namely the gravitational constant,
which we have set to unity. This in turn fixes the unit of time
and allows us to turn the numerical values 
for the period of the oscillations into values with proper units.
The diameter of elliptical galaxies, which can be close to spherically 
symmetric and are thus of interest here, ranges between
0.1 kpc and 100 kpc and their mass between $10^7$ and $10^{13}$ 
solar masses. The oscillation period predicted 
from a polytropic model with $k=1, l=0$ and $y(0)=0.6$
ranges between $1.7 \cdot 10^7$~years for ``small'' galaxies
and $5.4 \cdot 10^8$~years for ``large'' ones; $k=1, l=0$ and $y(0)=1.6$
yields $1.1 \cdot 10^7$~years for ``small'' and
$3.4 \cdot 10^8$~years for ``large'' galaxies.
Of course these numbers change with the steady state model, but we found that
the time scale of $10^7 - 10^8$~years seems typical, and it seems to be of
an astrophysically reasonable order of magnitude.

The analogous oscillations for
the Euler-Poisson system were used to explain for example
the Cepheid variables, cf.\ \cite{edd}. It should also be noted
that many arguments in the astrophysical analysis of galaxies
explicitly or implicitly rely on the assumption that they are
in some equilibrium. It seems debatable how justified 
this assumption is. In any case, the question whether galaxies can oscillate
has been asked in the astrophysics literature, cf.\ \cite{louis},
and if one models galaxies by the Vlasov-Poisson system,
then the answer to this question should be an emphatic ``Yes''.

%Close to a particular stable 
%steady state there seems to be a plethora of periodically oscillating
%solutions which are not damped and which can be triggered by a variety
%of perturbations. But we also saw that not all these perturbations
%lead to solutions which are exactly periodic from the start. There
%often seems to be an initial phase where the perturbed solution
%settles down to one which seems exactly time periodic. 
%The mechanism which drives the perturbed solutions to a close by nearly
%time periodic one could in principle be again of Landau damping nature.
%But to prove this, if true, is certainly even more challenging
%than the existence of the time periodic oscillations which
%we have observed numerically.
%
%

\end{document}